\documentclass[aps,superscriptaddress,eqsecnum,nofootinbib,preprintnumbers]{revtex4}
\usepackage{graphicx,epsfig}
\usepackage{amsmath}
\usepackage {amssymb}
\usepackage{subfigure}

\usepackage{relsize}
\newcommand{\be}{\begin{equation}}
\newcommand{\ee}{\end{equation}}
\newcommand{\bea}{\begin{eqnarray}}
\newcommand{\eea}{\end{eqnarray}}
\newcommand{\nn}{\nonumber}

\begin{document}

\title{On the action of the complete Brans-Dicke theories}

\author{Georgios Kofinas}
\email{gkofinas@aegean.gr} \affiliation{Research Group of Geometry,
Dynamical Systems and Cosmology,
Department of Information and Communication Systems Engineering\\
University of the Aegean, Karlovassi 83200, Samos, Greece}

\author{Minas Tsoukalas}
\email{minasts@central.ntua.gr} \affiliation{Physics Department, Bo\u{g}azi\c{c}i University, \\
34342, Bebek, Istanbul, Turkey}\affiliation{Physics Division, National Technical
University of Athens, 15780 Zografou Campus, Athens, Greece}

\begin{abstract}

Recently the most general completion of Brans-Dicke theory was appeared with energy exchanged
between the scalar field and ordinary matter, given that the equation
of motion for the scalar field keeps the simple wave form of Brans-Dicke.
This class of theories contain undetermined functions, but there exist only three theories
which are unambiguously determined from consistency. Here, for the first such theory, it is found
the action of the vacuum theory which arises as the limit of the full matter theory.
A symmetry transformation of this vacuum action in the Jordan frame is found which consists of a
conformal transformation of the metric together with a redefinition of the scalar field.
Since the general family of vacuum theories is parametrized by an arbitrary function of the scalar
field, the action of this family is also found. As for the full matter theory it is only
found the action of the system when the matter Lagrangian vanishes on-shell, as for example
for pressureless dust. Due to the interaction, this matter Lagrangian is non-minimally coupled
either in the Jordan or the Einstein frame.

\end{abstract}

\maketitle

\section{Introduction} \label{Introduction}

Scalar-tensor gravitational theories are studied extensively as an alternative to General
Relativity. Brans-Dicke theory \cite{Brans:1961sx} is a simple such theory which was initially
formulated in terms of an action constructed
from a metric $g_{\mu\nu}$ and a scalar field $\phi$, solely based on dimensional arguments, and with
the matter Lagrangian being minimally coupled. The effective gravitational constant of the theory
varies as the inverse of the scalar field, $G\sim\frac{1}{\phi}$, and there is no dimensionfull
parameter in the vacuum theory. The theory respects Mach's principle and weak equivalence principle.
In modern context Brans-Dicke theory appears naturally from supergravity models, from string theories
at low energies and from dimensional reduction of Kaluza-Klein theories \cite{Freund:1982pg}.
An alternative way to derive Brans-Dicke theory is to construct directly the field equations of motion
\cite{Weinberg} respecting the simple scalar field equation $\Box\phi=4\pi\lambda\mathcal{T}$, where
$\mathcal{T}=\mathcal{T}^{\mu}_{\,\,\,\,\mu}$ is the trace of the matter energy-momentum tensor
$\mathcal{T}^{\mu}_{\,\,\,\,\nu}$ and
$\lambda$ is a dimensionless coupling. The demand for this derivation is that the energy-momentum
tensor of the scalar field is made out of terms each of which involves two derivatives of one or two
$\phi$ fields, and $\phi$ itself. The theory gives the correct Newtonian weak-field limit and in order
to avoid the propagation of the fifth force, the coupling between matter and the massless field
$\phi$ should be suppressed, so $\lambda$ should be very small.
Recently there is an increasing interest in cosmology in
interacting models between dark matter and dark energy and such a mechanism can be useful to solve the
coincidence problem \cite{Amendola:1999qq}. However, usually such interactions are chosen ad-hoc
and do not arise by any physical theory.
In \cite{Bertolami:2007zm} it was actually argued that
observational evidence supports an interaction between dark matter and dark energy and violation
of equivalence principle between baryons and dark matter. In any case it would be interesting to
violate the standard conservation equation of $\mathcal{T}^{\mu}_{\,\,\,\,\nu}$ of Brans-Dicke theory.
For example, in \cite{Clifton:2006vm} an energy exchange model with a modified wave equation for
$\phi$ was considered (for other approaches with modified equations of motion see \cite{Smalley:1974gn}).
Useful piece of information and exhaustive analysis of Brans-Dicke gravity can be found in
\cite{maeda book}, \cite{faraoni book}, \cite{Clifton:2011jh}.
We should note that if the interaction model is to be worked out at the level of an action, then there
are various interactions of the matter Lagrangian with the scalar field, all having as limit the
Brans-Dicke action in the absence of interactions. The number of such actions can increase in the
presence of Newton's constant $G_{\!N}$ or new massive or massless parameters.
In \cite{Kofinas:2015nwa}, analyzing exhaustively the Bianchi identities, it was found the
general class of consistent theories generalizing Brans-Dicke theory, when the exact energy
conservation of the matter stress tensor was relaxed, while preserving the equation
$\Box\phi=4\pi\lambda\mathcal{T}$. This class of theories is parametrized by one or two free
functions of the scalar field, but it was found that there are only three theories, each
with a specific interaction term, which are unambiguously determined from consistency. These
unique and natural theories are certainly the predominant completions of Brans-Dicke theory.
In the present paper we are going to focus on the first such theory whose equations of motion
appear in section \ref{jefr} and they contain a new dimensionfull parameter
$\nu$ (which is actually an integration constant). In order for the equivalence
principle not to be violated at the ranges it has been tested, the parameters $\lambda,\nu$
should be chosen appropriately, and mechanisms such as Chameleon \cite{Khoury:2003aq} or Vainshtein
(self screening) \cite{Vainshtein:1972sx}, or even the existence of distinct conservation laws
for baryonic and non-baryonic matter could contribute to this direction.
Here, we will find in section \ref{mefr} the action of the vacuum part of this theory,
study its symmetry transformation in section \ref{meff}, and only partially answer the question
of its total action in section \ref{ejrr}. Moreover, in \cite{Kofinas:2015nwa} the general family
of the vacuum Brans-Dicke type of theories were found which satisfy the free wave equation for
the scalar field, and this family is parametrized by one free function of $\phi$. Here, the action of
these vacuum theories will be also found in section \ref{jwrf} and turns out to be a particular
sector of Horndeski family.

\section{Complete Brans-Dicke equations}
\label{jefr}

We start with the complete Brans-Dicke theory presented in \cite{Kofinas:2015nwa}
\begin{eqnarray}
&&\!\!\!\!\!\!\!G^{\mu}_{\,\,\,\nu}\!=\!\frac{8\pi}{\phi}
(T^{\mu}_{\,\,\,\nu}+\mathcal{T}^{\mu}_{\,\,\,\,\nu})
\label{elp}\\
&&\!\!\!\!\!\!\!T^{\mu}_{\,\,\,\nu}\!=\!\frac{\phi}{2\lambda(\nu\!+\!8\pi\phi^{2})^{2}}\Big{\{}
2\big[(1\!+\!\lambda)\nu\!+\!4\pi(2\!-\!3\lambda)\phi^{2}\big]\phi^{;\mu}\phi_{;\nu}
\!-\!\big[(1\!+\!2\lambda)\nu\!+\!4\pi(2\!-\!3\lambda)\phi^{2}\big]\delta^{\mu}_{\,\,\,\nu}
\phi^{;\rho}\phi_{;\rho} \Big{\}}
\!+\!\frac{\phi^{2}}{\nu\!+\!8\pi\phi^{2}}
\big(\phi^{;\mu}_{\,\,\,\,;\nu}\!-\!\delta^{\mu}_{\,\,\,\nu}\Box\phi\big)
\nn\\
\label{utd}\\
&&\!\!\!\!\!\!\!\Box\phi\!=\!4\pi\lambda\mathcal{T}\label{lrs}\\
&&\!\!\!\!\!\!\!\mathcal{T}^{\mu}_{\,\,\,\,\nu;\mu}\!=\!\frac{\nu}{\phi(\nu\!+\!8\pi\phi^{2})}
\mathcal{T}^{\mu}_{\,\,\,\,\nu}\phi_{;\mu}\,.
\label{idj}
\end{eqnarray}
The parameter $\nu$ is arbitrary and arises as an integration constant from the integration procedure
(its dimensions are mass to the fourth). The parameter $\lambda\neq 0$ is related
to the standard Brans-Dicke parameter $\omega_{BD}=\frac{2-3\lambda}{2\lambda}$ and controls the
strength of the interaction in (\ref{lrs}), while $\nu$ controls the strength of the interaction
in (\ref{idj}). This theory arises out of consistency given that the scalar field equation of
motion is (\ref{lrs}), and $T^{\mu}_{\,\,\,\,\nu}$ is constructed from terms each of which involves
two derivatives of one or two $\phi$ fields and $\phi$ itself.
The right-hand side of equation (\ref{elp}) is consistent with
the Bianchi identities, i.e. it is covariantly conserved on-shell, and therefore, the system
of equations (\ref{elp})-(\ref{idj}) is well-defined.
Moreover, it is the unique theory with an interaction term of the form
$\mathcal{T}^{\mu}_{\,\,\,\,\nu;\mu}\sim\mathcal{T}^{\mu}_{\,\,\,\,\nu}\phi_{;\mu}$.
For $\nu=0$ it reduces to the Brans-Dicke theory \cite{Brans:1961sx} (in units
with $c=1$)
\begin{eqnarray}
G^{\mu}_{\,\,\,\nu}\!\!&=&\!\!\frac{8\pi}{\phi}(T^{\mu}_{\,\,\,\nu}+\mathcal{T}^{\mu}_{\,\,\,\,\nu})
\label{kns}\\
T^{\mu}_{\,\,\,\nu}\!\!&=&\!\!\frac{2-3\lambda}{16\pi\lambda\phi}\Big(
\phi^{;\mu}\phi_{;\nu}\!-\!\frac{1}{2}\delta^{\mu}_{\,\,\,\nu}
\phi^{;\rho}\phi_{;\rho} \Big)
\!+\!\frac{1}{8\pi}\big(\phi^{;\mu}_{\,\,\,\,;\nu}\!-\!\delta^{\mu}_{\,\,\,\nu}\Box\phi\big)
\label{qqd}\\
\Box\phi\!\!&=&\!\!4\pi\lambda\mathcal{T}\label{lwn}\\
\,\,\mathcal{T}^{\mu}_{\,\,\,\,\nu;\mu}\!\!&=&\!\!0\,.
\label{jrk}
\end{eqnarray}
The role of the new parameter $\nu$ is manifest in (\ref{idj}) and measures the deviation from the
exact conservation of matter.
The Lagrangian of the Brans-Dicke theory is
\begin{equation}
S_{BD}=\frac{1}{16\pi}\int \!d^{4}x \,\sqrt{-g} \,\Big(\phi R-\frac{\omega_{BD}}{\phi}
g^{\mu\nu}\phi_{,\mu}\phi_{,\nu}\Big)+\int \!d^{4}x \,\sqrt{-g} \,L_{m}\,,
\label{jsw}
\end{equation}
where $L_{m}(g_{\kappa\lambda},\Psi)$ is the matter Lagrangian
depending on some extra fields $\Psi$.

\section{The Vacuum Lagrangian}
\label{mefr}

We will find here the action of the vacuum theory arising by setting $\mathcal{T}^{\mu}_{\,\,\,\nu}$
to zero in the above theory. Mimicking the action (\ref{jsw}), we consider an action of the form
\begin{equation}
S_{g}=\frac{1}{16\pi}\int\!d^{4}x\,\sqrt{-g}\,\big[f(\phi)R-h(\phi)\phi^{;\mu}\phi_{;\mu}\big]\,,
\label{ker}
\end{equation}
and we are looking to see if there are functions $f,h$ such that equations (\ref{elp})-(\ref{idj})
with $\mathcal{T}^{\mu}_{\,\,\,\,\nu}=0$ arise under variation of (\ref{ker}). Action (\ref{ker})
is a sector of the Horndeski Lagrangian \cite{Horndeski:1974wa}, \cite{Deffayet:2009mn} which leads to
the most general field equations with second order derivatives. Hopefully, the Lagrangian (\ref{ker})
will be enough for our purposes. Variation of (\ref{ker}) with respect to the metric gives, up to
boundary terms
\begin{equation}
\delta_{g}S_{g}=-\frac{1}{16\pi}\int\!d^{4}x\,\sqrt{-g}\,
\Big[fG^{\mu\nu}-(f''\!+\!h)\phi^{;\mu}\phi^{;\nu}+\Big(f''\!+\!\frac{1}{2}h\Big)g^{\mu\nu}\phi^{;\rho}
\phi_{;\rho}-f'\big(\phi^{;\mu;\nu}\!-\!g^{\mu\nu}\Box\phi\big)\Big]
\delta g_{\mu\nu}\,,
\label{hef}
\end{equation}
where a prime denotes differentiation with respect to $\phi$ and a $;$ stands for the covariant
differentiation with respect to $g_{\mu\nu}$. Therefore, the gravitational field equation is
\begin{equation}
\mathcal{E}^{\mu}_{\,\,\,\nu}\equiv G^{\mu}_{\,\,\,\nu}-
\frac{1}{f}(f''\!+\!h)\phi^{;\mu}\phi_{;\nu}+\frac{1}{f}\Big(f''\!+\!\frac{1}{2}h\Big)
\delta^{\mu}_{\,\,\,\nu}\phi^{;\rho}\phi_{;\rho}-
\frac{f'}{f}\big(\phi^{;\mu}_{\,\,\,\,;\nu}\!-\!\delta^{\mu}_{\,\,\,\nu}\Box\phi\big)=0\,.
\label{fhw}
\end{equation}
The trace of equations (\ref{fhw}) gives
\begin{equation}
\mathcal{E}^{\mu}_{\,\,\,\mu}=-R+\frac{1}{f}(3f''\!+\!h)\phi^{;\mu}\phi_{;\mu}+3\frac{f'}{f}\Box\phi=0\,.
\label{wem}
\end{equation}
In order for (\ref{fhw}) to coincide with equation (\ref{elp}), the following conditions on the
functions $f,h$ should be satisfied
\begin{eqnarray}
&&\frac{f'}{f}=\frac{8\pi\phi}{\nu\!+\!8\pi\phi^{2}}\label{jqw}\\
&&\frac{f''}{f}+\frac{h}{f}=\frac{8\pi}{\lambda(\nu\!+\!8\pi\phi^{2})^{2}}
\big[(1\!+\!\lambda)\nu\!+\!4\pi(2\!-\!3\lambda)\phi^{2}\big]\label{wrv}\\
&&\frac{f''}{f}+\frac{1}{2}\,\frac{h}{f}=\frac{4\pi}{\lambda(\nu\!+\!8\pi\phi^{2})^{2}}
\big[(1\!+\!2\lambda)\nu\!+\!4\pi(2\!-\!3\lambda)\phi^{2}\big]\,.\label{wrw}
\end{eqnarray}
Although equations (\ref{jqw})-(\ref{wrw}) form a system of three conditions for the two unknowns
$f,h$, it is however consistent. Indeed, differentiating (\ref{jqw}) with respect to $\phi$
and combining with (\ref{wrv}) we get
\begin{equation}
\frac{h}{f}=\frac{8\pi}{\lambda(\nu\!+\!8\pi\phi^{2})^{2}}
\big[\nu\!+\!4\pi(2\!-\!3\lambda)\phi^{2}\big]\,.
\label{wlo}
\end{equation}
Also, subtracting equations (\ref{wrv}), (\ref{wrw}) we obtain again (\ref{wlo}). Thus,
we are left with the system of the two equations (\ref{jqw}), (\ref{wlo}).
The solution of this system is
\begin{eqnarray}
&&f=\texttt{c} \sqrt{|\nu\!+\!8\pi\phi^{2}|}\label{lhf}\\
&&h= \texttt{c} \frac{8\pi}{\lambda}\,
\frac{\nu\!+\!4\pi(2\!-\!3\lambda)\phi^{2}}{|\nu\!+\!8\pi\phi^{2}|^{3/2}}\,,
\label{wof}
\end{eqnarray}
where $\texttt{c}$ is integration constant.

What remains is the satisfaction of equation (\ref{lrs}), namely $\Box\phi=0$.
The variation of (\ref{ker}) with respect to the scalar field gives, up to boundary terms
\begin{equation}
\delta_{\phi}S_{g}=\frac{1}{16\pi}\int\!d^{4}x\,\sqrt{-g}\,\big(f'R+h'\phi^{;\mu}\phi_{;\mu}
+2h\Box\phi\big)\delta\phi\,.\label{qwk}
\end{equation}
The scalar field equation is
\begin{equation}
\mathcal{E}_{\phi}\equiv f'R+h'\phi^{;\mu}\phi_{;\mu}+2h\Box\phi=0\,.
\label{owr}
\end{equation}
Using (\ref{wem}) to substitute $R$ in (\ref{owr}) we obtain
\begin{equation}
\Big(3\frac{f'^{2}}{f}\!+\!2h\Big)\Box\phi+\Big[f'\Big(3\frac{f''}{f}\!+\!\frac{h}{f}\Big)\!+\!h'\Big]
\phi^{;\mu}\phi_{;\mu}=0\,.
\label{lwf}
\end{equation}
Using equations (\ref{wrv}), (\ref{wlo}) to get the quantity $\frac{f''}{f}$, and also the solution
(\ref{lhf}), (\ref{wof}), we find that the coefficient of $\phi^{;\mu}\phi_{;\mu}$ in (\ref{lwf})
vanishes. Therefore, the scalar field equation (\ref{lwf}) becomes
\begin{equation}
\frac{16\pi\epsilon\texttt{c}}{\lambda\sqrt{|\nu\!+\!8\pi\phi^{2}|}}\Box\phi=0\,,
\label{srf}
\end{equation}
where $\epsilon=\text{sgn}(\nu\!+\!8\pi\phi^{2})$, which means $\Box\phi=0$. When $\epsilon>0$,
it is either $\nu>0$ or $\nu<0$, $|\phi|>\sqrt{\frac{|\nu|}{8\pi}}$. When $\epsilon<0$,
it is $\nu<0$, $|\phi|<\sqrt{\frac{|\nu|}{8\pi}}$.

Finally, we choose the integration constant $\texttt{c}=\frac{\eta}{\sqrt{8\pi}}$, where
$\eta=\text{sgn}(\phi)$,  to normalize the
action (\ref{ker}) to the Brans-Dicke action (\ref{jsw}) in the limit $\nu=0$.
The result is that the vacuum system (\ref{elp})-({\ref{idj}}) admits a Lagrangian and its action is
\begin{equation}
S_{g}=\frac{\eta}{2(8\pi)^{3/2}}\int\!d^{4}x\,\sqrt{-g}\,
\Big[\sqrt{|\nu\!+\!8\pi\phi^{2}|}\,R-\frac{8\pi}{\lambda}\,
\frac{\nu\!+\!4\pi(2\!-\!3\lambda)\phi^{2}}{|\nu\!+\!8\pi\phi^{2}|^{3/2}}
g^{\mu\nu}\phi_{,\mu}\phi_{,\nu}\Big]\,.
\label{jgk}
\end{equation}

\section{Symmetry transformation of the vacuum action}
\label{meff}

In this section we will find a transformation of the fields $(g_{\mu\nu},\phi)\rightarrow
(\hat{g}_{\mu\nu},\chi)$ such that the vacuum action (\ref{jgk}) remains form invariant, i.e. it is
written as
\begin{equation}
S_{g}=\frac{\eta}{2(8\pi)^{3/2}}\int\!d^{4}x\,\sqrt{-\hat{g}}\,
\Big[\sqrt{|\nu\!+\!8\pi\chi^{2}|}\,\hat{R}-\frac{8\pi}{\lambda}\,
\frac{\nu\!+\!4\pi(2\!-\!3\lambda)\chi^{2}}{|\nu\!+\!8\pi\chi^{2}|^{3/2}}
\hat{g}^{\mu\nu}\chi_{,\mu}\chi_{,\nu}\Big]\,.
\label{jgm}
\end{equation}
This transformation will therefore be a symmetry of the vacuum action in the Jordan frame.
To be precise, we consider a conformal transformation of $g_{\mu\nu}$ together with a field
redefinition for $\phi$, namely
\begin{equation}
\hat{g}_{\mu\nu}=\hat{\Omega}^{2}g_{\mu\nu}\,\,\,\,\,\,\,\,\,\,,\,\,\,\,\,\,\,\,\,\,\phi=A(\chi)\,.
\label{bel}
\end{equation}
If $\hat{R}$, $\hat{\Box}$ correspond to $\hat{g}_{\mu\nu}$, we have the relation
\begin{equation}
R=\hat{\omega}^{-2}\Big(\hat{R}-6\frac{\hat{\Box}\hat{\omega}}{\hat{\omega}}\Big)\,,
\label{wrg}
\end{equation}
where $\hat{\omega}=\hat{\Omega}^{-1}$.
Then, since $\hat{\Box}\hat{\omega}=\hat{\omega}''\phi^{|\mu}\phi_{|\mu}+\hat{\omega}'\,\hat{\Box}\phi$,
the action $S_{g}$ takes the form
\begin{equation}
S_{g}=\frac{1}{16\pi}\int\!d^{4}x\,\sqrt{-\hat{g}}\,f\hat{\omega}^{2}\Big[\hat{R}
-\Big(6\frac{\hat{\omega}''}{\hat{\omega}}\!+\!\frac{h}{f}\Big)\phi^{|\mu}\phi_{|\mu}
-6\frac{\hat{\omega}'}{\hat{\omega}}\hat{\Box}{\phi}\Big]\,,
\label{mgd}
\end{equation}
where a $|$ denotes covariant differentiation with respect to $\hat{g}_{\mu\nu}$
and a prime denotes as usual a differentiation with respect to $\phi$.
After an integration by parts, equation (\ref{mgd}) becomes, up to boundary terms
\begin{equation}
S_{g}=\frac{1}{16\pi}\int\!d^{4}x\,\sqrt{-\hat{g}}\,f\hat{\omega}^{2}
\Big[\hat{R}+\Big(6\frac{f'\hat{\omega}'}{f\hat{\omega}}\!+\!6\frac{\hat{\omega}'^{2}}
{\hat{\omega}^{2}}\!-\!\frac{h}{f}\Big)\phi^{|\mu}\phi_{|\mu}\Big]\,,
\label{lem}
\end{equation}
and finally
\begin{equation}
S_{g}=\frac{1}{16\pi}\int\!d^{4}x\,\sqrt{-\hat{g}}\,f\hat{\omega}^{2}
\Big[\hat{R}+\Big(6\frac{f'\hat{\omega}'}{f\hat{\omega}}\!+\!6\frac{\hat{\omega}'^{2}}
{\hat{\omega}^{2}}\!-\!\frac{h}{f}\Big)\Big(\frac{dA}{d\chi}\Big)^{\!2}\,
\hat{g}^{\mu\nu}\chi_{,\mu}\chi_{\nu}\Big]\,.
\label{ltg}
\end{equation}
Action (\ref{ltg}) is also written as
\begin{equation}
S_{g}=\frac{1}{16\pi}\int\!d^{4}x\,\sqrt{-\hat{g}}\,f\hat{\omega}^{2}
\Big{\{}\hat{R}+\Big[6\frac{\hat{\omega}'}{\hat{\omega}}
\,\frac{(f\hat{\omega}^{2})'}{f\hat{\omega}^{2}}\!-\!6\frac{\hat{\omega}'^{2}}
{\hat{\omega}^{2}}\!-\!\frac{h}{f}\Big]\Big(\frac{dA}{d\chi}\Big)^{\!2}\,
\hat{g}^{\mu\nu}\chi_{,\mu}\chi_{\nu}\Big{\}}\,.
\label{mje}
\end{equation}
In order for (\ref{mje}) to be identified with (\ref{jgm}) it should be
\begin{eqnarray}
&&f\hat{\omega}^{2}=\hat{f}\label{klp}\\
&&\Big[6\frac{\hat{\omega}'}{\hat{\omega}}
\Big(\frac{\hat{f}'}{\hat{f}}\!-\!\frac{\hat{\omega}'}{\hat{\omega}}\Big)
\!-\!\frac{h}{f}\Big]\Big(\frac{dA}{d\chi}\Big)^{\!2}=-\frac{\hat{h}}{\hat{f}}\,,
\label{kie}
\end{eqnarray}
where
\begin{equation}
\hat{f}=\frac{\eta}{\sqrt{8\pi}} \sqrt{|\nu\!+\!8\pi\chi^{2}|}
\,\,\,\,\,\,\,\,\,\,,\,\,\,\,\,\,\,\,\,\,
\hat{h}=\frac{\eta\sqrt{8\pi}}{\lambda} \,
\frac{\nu\!+\!4\pi(2\!-\!3\lambda)\chi^{2}}{|\nu\!+\!8\pi\chi^{2}|^{3/2}}\,.
\label{low}
\end{equation}
Converting the $\phi-$derivatives in (\ref{kie}) into $\chi-$derivatives we get
\begin{equation}
\frac{6}{\hat{\omega}}\,\frac{d\hat{\omega}}{d\chi}\Big(\frac{1}{\hat{f}}\,
\frac{d\hat{f}}{d\chi}-\frac{1}{\hat{\omega}}\,\frac{d\hat{\omega}}{d\chi}\Big)
-\frac{h}{f}\Big(\frac{dA}{d\chi}\Big)^{\!2}=-\frac{\hat{h}}{\hat{f}}\,.
\label{plm}
\end{equation}
Since $\frac{2}{\hat{\omega}}\frac{d\hat{\omega}}{d\chi}=\frac{1}{\hat{\omega}^{2}}
\frac{d(\hat{\omega}^{2})}{d\chi}$, we have from (\ref{klp})
\begin{equation}
\frac{2}{\hat{\omega}}\frac{d\hat{\omega}}{d\chi}=\frac{1}{\hat{f}}
\frac{d\hat{f}}{d\chi}-\frac{1}{f}\frac{df}{d\phi}\frac{dA}{d\chi}\,.
\label{vbr}
\end{equation}
Substituting (\ref{vbr}) into (\ref{plm}) we obtain
\begin{equation}
\Big[\Big(\frac{1}{f}\,\frac{df}{d\phi}\Big)^{\!2}\!+\!\frac{2h}{3f}\Big]\Big(\frac{dA}{d\chi}\Big)
^{\!2}=\Big(\frac{1}{\hat{f}}\,\frac{d\hat{f}}{d\chi}\Big)^{\!2}\!+\!\frac{2\hat{h}}{3\hat{f}}\,,
\label{yth}
\end{equation}
which furthermore gets a separable form
\begin{equation}
\frac{d\phi}{\sqrt{|\nu\!+\!8\pi\phi^{2}|}}=\pm \frac{d\chi}{\sqrt{|\nu\!+\!8\pi\chi^{2}|}}\,,
\label{lmy}
\end{equation}
where $\text{sgn}(\nu\!+\!8\pi\chi^{2})=\epsilon$.

For $\epsilon>0$, integration of (\ref{lmy}) gives
\begin{equation}
\phi=\frac{s}{8\pi}\Big(\theta\big|4\pi\chi\!+\!\sqrt{2\pi}\sqrt{\nu\!+\!8\pi\chi^{2}}\big|^{\pm 1}
-\frac{2\pi\nu}{\theta}\big|4\pi\chi\!+\!\sqrt{2\pi}\sqrt{\nu\!+\!8\pi\chi^{2}}\big|^{\mp 1}\Big)\,,
\label{hyv}
\end{equation}
where $\theta>0$ is integration constant and $s=\text{sgn}\big(4\pi\phi+\sqrt{2\pi}
\sqrt{\nu\!+\!8\pi\phi^{2}}\big)=
\text{sgn}\big(\theta\big|4\pi\chi\!+\!\sqrt{2\pi}\sqrt{\nu\!+\!8\pi\chi^{2}}\big|^{\pm 1}
+\frac{2\pi\nu}{\theta}
\big|4\pi\chi\!+\!\sqrt{2\pi}\sqrt{\nu\!+\!8\pi\chi^{2}}\big|^{\mp 1}\big)$, or inversely
\begin{equation}
\chi=\frac{s'}{8\pi}\Big(\theta'\big|4\pi\phi\!+\!\sqrt{2\pi}\sqrt{\nu\!+\!8\pi\phi^{2}}\big|^{\pm 1}
-\frac{2\pi\nu}{\theta'}\big|4\pi\phi\!+\!\sqrt{2\pi}\sqrt{\nu\!+\!8\pi\phi^{2}}\big|^{\mp 1}\Big)\,,
\label{hyb}
\end{equation}
where $\theta'=\theta^{\mp 1}>0$ and $s'=\text{sgn}\big(4\pi\chi+\sqrt{2\pi}
\sqrt{\nu\!+\!8\pi\chi^{2}}\big)=
\text{sgn}\big(\theta'\big|4\pi\phi\!+\!\sqrt{2\pi}\sqrt{\nu\!+\!8\pi\phi^{2}}\big|^{\pm 1}
+\frac{2\pi\nu}{\theta'}\big|4\pi\phi\!+\!\sqrt{2\pi}\sqrt{\nu\!+\!8\pi\phi^{2}}\big|^{\mp 1}\big)$.

For $\epsilon<0$ it is
\begin{equation}
\phi=\sqrt{\frac{|\nu|}{8\pi}}\,\sin\Big[c_{1}\!\pm\!\arcsin\Big(\sqrt{\frac{8\pi}{|\nu|}}\,\chi\Big)
\Big]\,,
\label{gyr}
\end{equation}
where $c_{1}$ is integration constant, or inversely
\begin{equation}
\chi=\pm\sqrt{\frac{|\nu|}{8\pi}}\,\sin\Big[\arcsin\Big(\sqrt{\frac{8\pi}{|\nu|}}\,\phi\Big)
\!-\!c_{1}\Big]\,.
\label{gyb}
\end{equation}

Finally, since $\phi(\chi)$ or $\chi(\phi)$ have been found, the conformal transformation (\ref{bel}),
which leaves the vacuum action $S_{g}$ form-invariant, is
\begin{equation}
\hat{g}_{\mu\nu}=\sqrt{\frac{|\nu\!+\!8\pi\phi^{2}|}{|\nu\!+\!8\pi\chi^{2}|}}\,\,g_{\mu\nu}\,.
\label{drl}
\end{equation}
From (\ref{hyv}), (\ref{drl}) we see that in the Brans-Dicke limit $\nu=0$, we get $\chi\propto
\phi^{-1}$, $\hat{g}_{\mu\nu}\propto \phi^{2}g_{\mu\nu}$, which leads to a symmetry transformation
of the Brans-Dicke action (\ref{jsw}) \cite{faraoni book}.

\section{The Lagrangian of general vacuum Brans-Dicke theories}
\label{jwrf}

Setting $\mathcal{T}^{\mu}_{\,\,\,\,\nu}=0$ in the system (\ref{elp})-(\ref{idj}) an extended
vacuum Brans-Dicke theory arises, which for $\nu=0$ reduces to the vacuum Brans-Dicke theory,
and this was studied in the previous sections.
However, it was shown in \cite{Kofinas:2015nwa} that this theory is not the most general vacuum
theory respecting the wave equation $\Box\phi=0$ and the standard assumption for
$T^{\mu}_{\,\,\,\nu}$
being a sum of terms with two derivatives. The most general such theory is
\begin{eqnarray}
&&\!\!\!\!\!\!\!G^{\mu}_{\,\,\,\nu}\!=\!\frac{8\pi}{\phi}T^{\mu}_{\,\,\,\nu}
\label{elf}\\
&&\!\!\!\!\!\!\!T^{\mu}_{\,\,\,\nu}\!=\!A(\phi)\phi^{;\mu}\phi_{;\nu}
\!+\!B(\phi)\delta^{\mu}_{\,\,\,\nu}\phi^{;\rho}\phi_{;\rho}\!+\!C(\phi)\phi^{;\mu}_{\,\,\,\,;\nu}
\label{urd}\\
&&\!\!\!\!\!\!\!\Box\phi\!=\!0\label{lts}\,,
\end{eqnarray}
where the coefficients $A,B,C$ satisfy the differential equations
\begin{eqnarray}
&&A'\!+\!B'\!+\!\frac{4\pi}{\phi}C(A\!-\!2B)\!-\!\frac{1}{\phi}(A\!+\!B)=0
\label{jdt}\\
&&C'\!+\!\frac{8\pi}{\phi}C^{2}\!-\!\frac{1}{\phi}C\!+\!A\!+\!2B=0\,.
\label{kdw}
\end{eqnarray}
The energy-momentum tensor $T^{\mu}_{\,\,\,\nu}$ of equation (\ref{utd}) is easily seen that
satisfies the system (\ref{jdt}), (\ref{kdw}) and defines probably one of the most interesting
vacuum theories. However, the solution of equations (\ref{jdt}), (\ref{kdw}) in principle contains one
arbitrary function of $\phi$. Here, we will show that the general vacuum theory defined by the system
(\ref{jdt}), (\ref{kdw}) arises from an action and we will find the general such action.
This also contains an arbitrary function and provides the field equations (\ref{elf})-(\ref{lts}).

We start with the Horndeski theory which consists of the most general Lagrangian \cite{Horndeski:1974wa}
providing second order field equations for both the metric and the scalar field.
This theory was recently rediscovered independently \cite{Deffayet:2009mn} and cast in a simpler
form, having the following structure
\begin{equation}
S=\int \!d^{4}x\,\sqrt{-g}\,\big(\mathcal{L}_{2}+\mathcal{L}_{3}+\mathcal{L}_{4}+\mathcal{L}_{5} \big),
\label{horn}
\end{equation}
where
\begin{eqnarray}
\mathcal{L}_{2}&=&G_{2}\\
\mathcal{L}_{3}&=&-G_{3}\square \phi\\
\mathcal{L}_{4}&=&G_{4}R+G_{4X}
\left[(\square\phi)^{2}-\phi_{;\mu;\nu}\phi^{;\mu;\nu} \right]\\
\mathcal{L}_{5}&=&G_{5}G_{\mu\nu}\phi^{;\mu;\nu}-\frac{1}{6}G_{5X}
\left[(\square \phi)^{3}+2\phi_{;\mu;\nu}\phi^{;\kappa;\mu}\phi_{;\kappa}^{\,\,\,\,\,;\nu}
-3\phi_{;\mu;\nu}\phi^{;\mu;\nu}\square\phi \right].
\end{eqnarray}
The functions $G_{i}$ ($i=2,3,4,5$) depend on the scalar field $\phi$ and its kinetic
energy $X=-\frac{1}{2}\phi^{;\mu}\phi_{;\mu}$, i.e. $G_{i}=G_{i}(\phi,X)$.
The field equations for the metric and the scalar field stemming from the variation of (\ref{horn})
are respectively the following \cite{Maselli:2015yva}
\begin{eqnarray}
\!\!\!\!\!\!\!\!\!\!
{\cal E}_{\mu\nu}&\!\!\!\!\!\!=\!\!\!\!\!\!&
-\frac{1}{2}G_2 g_{\mu\nu}+G_{2X}X_{\mu\nu}-\Big[G_{3X}X_{\mu\nu}\square\phi+
\frac{1}{2}g_{\mu\nu}G_{3\kappa}\phi^{\kappa}-G_{3(\mu}\phi_{\nu)}\Big]\nn\\
&&+G_4 G_{\mu\nu}+RG_{4X}X_{\mu\nu}+G_{4\kappa}{^{\kappa}}g_{\mu\nu}-G_{4\mu\nu}+
\Big(G_{4XX}X_{\mu\nu}-\frac{1}{2}G_{4X}g_{\mu\nu}\Big)
\big[(\square\phi)^2-\phi_{\mu\nu}^2\big]+2G_{4X}
\phi_{\mu\nu}\square\phi\nn\\
&&-2[G_{4X}\phi_{;(\mu}\square\phi]_{;\nu)}+(G_{4X}\phi^{\kappa}\square\phi)_{;\kappa}g_{\mu\nu}+
2[G_{4X}\phi_{(\mu}\phi^{\kappa}{_{\nu)}}]_{;\kappa}-(G_{4X}\phi^{\kappa}\phi_{\mu\nu})_{;\kappa}
-2G_{4X}\phi_{\nu\kappa}\phi^{\kappa}{_{\mu}}\nn\\
&&+{G}_{\kappa\lambda}\phi^{\kappa\lambda}
\Big(G_{5X}X_{\mu\nu}-\frac{1}{2}G_5g_{\mu\nu}\Big)+2G_5\phi^{\kappa}{_{(\nu}}{G}_{\mu)\kappa}-
[G_{5}\phi_{(\mu}{G}_{\nu)\kappa}]^{;\kappa}+\frac{1}{2}(G_{5}\phi_{\kappa}{G}_{\mu\nu})^{;\kappa}
+\frac{1}{2}\Big\{RG_5\phi_{\mu\nu}\nn\\
&&-G_5\phi{_{\kappa}}^{\kappa}R_{\mu\nu}
+\square(G_5\phi_{\mu\nu})+(G_5\phi{_{\kappa}}^{\kappa})_{;\nu;\mu}
-2[G_5\phi{_{(\mu}}^{\kappa}]_{;\nu);\kappa}+
\big[(G_5\phi^{\kappa\lambda})_{;\lambda;\kappa}
-\square(G_5\phi{_{\kappa}}^{\kappa})\big]g_{\mu\nu}\Big\}\nn\\
&&-\frac{1}{6}\Big(G_{5XX}X_{\mu\nu}-\frac{1}{2}G_{5X}g_{\mu\nu}\Big)
\big[(\square\phi)^3+2\phi_{\kappa\lambda}^3-3\phi_{\kappa\lambda}^2\square\phi\big]
-\frac{1}{2}\Big\{G_{5X}(\square\phi)^2\phi_{\mu\nu}
-2[G_{5X}(\square\phi)^2\phi_{(\mu}]_{;\nu)}\nn\\
&&+\frac{1}{2}\big[G_{5X}(\square\phi)^2\phi_{\kappa}\big]^{;\kappa}g_{\mu\nu}\Big\}
-\Big\{G_{5X}\phi_{\kappa\mu}\phi_{\nu\lambda}\phi^{\lambda\kappa}
-\big[G_{5X}\phi_{(\mu}\phi^{\lambda}{_{\nu)}}\phi_{\lambda\kappa}\big]^{;\kappa}+
\frac{1}{2}\big(G_{5X}\phi_{\kappa}\phi_{\lambda\mu}\phi^{\lambda}{_{\nu}}\big)^{;\kappa}\Big\}\nn\\
&&+\frac{1}{2}\Big\{G_{5X}\big(\phi_{\kappa\lambda}^2\phi_{\mu\nu}+2\phi_{\mu\kappa}
\phi^{\kappa}{_{\nu}}\square\phi\big)
-[G_{5X}\phi_{\kappa\lambda}\phi^{\kappa\lambda}\phi_{(\mu}]_{;\nu)}
+\frac{1}{2}\big(G_{5X}\phi_{\kappa}\phi_{\lambda\rho}\phi^{\lambda\rho}\big)^{;\kappa}g_{\mu\nu}\nn\\
&&-2\big[G_{5X}\phi_{(\mu}\phi_{\nu)\kappa}\square\phi\big]^{;\kappa}+
(G_{5X}\phi_{\kappa}\phi_{\mu\nu}\square\phi)^{;\kappa}\Big\}=0
\label{eab}
\end{eqnarray}
\begin{eqnarray}
\!\!\!\!\!\!\!\!\!\!\!\!\!
{\cal E}_{\phi}&\!\!=\!\!\!\!\!\!&\phantom{+}G_{2\phi}+(G_{2X}\phi^{\mu})_{;\mu}
-\big[G{_{3\mu}}^{\mu}+(G_{3X}\phi^{\mu}\square\phi)_{;\mu}
+G_{3\phi}\square\phi\big]\nn\\
&&+RG_{4\phi}+G_{4X\phi}[(\square\phi)^2-\phi_{\mu\nu}^2]
+\big{\{}G_{4XX}\phi_{\kappa}[(\square\phi)^2-\phi_{\mu\nu}^2]\big{\}}^{;\kappa}
+(RG_{4X}\phi_{\mu})^{;\mu}
+2\square(G_{4X}\square\phi)-2(G_{4X}\phi_{\mu\nu})^{;\nu;\mu}\nn\\
&&+G_{5\phi}{G}_{\mu\nu}\phi^{\mu\nu}
+G_{5}^{\mu\nu}{G}_{\mu\nu}-\frac{1}{6}G_{5X\phi}\big[(\square\phi)^3+2\phi_{\mu\nu}^3
-3\phi_{\mu\nu}^2\square\phi\big]+(G_{5X}G^{\mu\nu}\phi_{\mu\nu}\phi_{\kappa})^{;\kappa}
-\frac{1}{2}\square\big[G_{5X}(\square\phi)^2\big]\nn\\
&&-\frac{1}{6}\big{\{}G_{5XX}\phi_{\kappa}\big[(\square\phi)^3
+2\phi_{\mu\nu}^3-3\phi_{\mu\nu}^2\square\phi\big]\big{\}}^{;\kappa}
-(G_{5X}\phi^{\kappa}{_{\mu}}\phi_{\kappa\nu})^{;\nu;\mu}
+\frac{1}{2}\square(G_{5X}\phi_{\mu\nu}^2)+
(G_{5X}\phi_{\mu\nu}\square\phi)^{;\nu;\mu}=0\,.
\label{ep}
\end{eqnarray}
A subscript $\phi$ or $X$ denotes a partial differentiation with respect to $\phi$ or $X$, while
$X_{\mu\nu}=\frac{\partial X}{\partial g^{\mu\nu}}=-\frac{1}{2}\phi_{,\mu}\phi_{,\nu}$.
Parentheses around a couple of indices mean symmetrization with the factor 1/2 included.
Also, we denote $f_{\mu\dots\nu}=f_{;\nu\dots \,;\mu}$ for a function $f$, while
$\phi_{\mu\nu}^2=\phi_{\mu\nu}\phi^{\mu\nu}$ and
$\phi_{\mu\nu}^3=\phi_{\mu\nu}\phi^{\nu\kappa}\phi^{\mu}{_{\kappa}}$.
Despite the fact of the appearance of higher derivatives in (\ref{eab}), (\ref{ep}), the field
equations can be reduced to second order using appropriate identities.

Equations (\ref{eab}), (\ref{ep}) should coincide with equations (\ref{elf}), (\ref{lts}).
All terms in (\ref{eab}), (\ref{ep}) multiplied by a $X$ derivative of $G_{5}$ contain
more than two derivatives on $\phi$. However, combining suitable such terms and using the formula
connecting two successive derivatives with the Riemann tensor, terms with lower number of derivatives
arise, which are multiplied by the Riemann tensor. Since there are no such structures in
(\ref{elf})-(\ref{lts}), all these terms should vanish. This requires that
$G_{5X}=0$, i.e. $G_{5}=G_{5}(\phi)$. Similarly, various terms in (\ref{eab}), (\ref{ep})
multiplied by a $X$ derivative of $G_{4}$ contain more than two derivatives on $\phi$ and therefore
$G_{4X}=0$, i.e. $G_{4}=G_{4}(\phi)$. The existence of quantities of the form
$G_{3\mu}\phi_{\nu}$ in (\ref{eab}) makes necessary that $G_{3X}=0$, i.e.
$G_{3}=G_{3}(\phi)$. Finally, the existence of the term $(G_{2X}\phi^{\mu})_{;\mu}$ in (\ref{ep}) leads
to $G_{2XX}=0$, i.e. $G_{2X}=G_{2X}(\phi)$, which means $G_{2}=\gamma_{2}(\phi)+g_{2}(\phi)X$.
Thus, equations (\ref{eab}), (\ref{ep}) take the form
\begin{eqnarray}
{\cal E}_{\mu\nu}&\!\!=\!\!&
-\frac{1}{2}G_2 g_{\mu\nu}+G_{2X}X_{\mu\nu}
+G_{3(\mu}\phi_{\nu)}-\frac{1}{2}g_{\mu\nu}G_{3\kappa}\phi^{\kappa}
+G_4 G_{\mu\nu}+G_{4\kappa}{^{\kappa}}g_{\mu\nu}-G_{4\mu\nu}\nn\\
&&-\frac{1}{2}G_5{G}_{\kappa\lambda}\phi^{\kappa\lambda}
g_{\mu\nu}+2G_5\phi^{\kappa}{_{(\nu}}{G}_{\mu)\kappa}-
[G_{5}\phi_{(\mu}{G}_{\nu)\kappa}]^{;\kappa}+\frac{1}{2}(G_{5}\phi_{\kappa}{G}_{\mu\nu})^{;\kappa}
+\frac{1}{2}\Big\{RG_5\phi_{\mu\nu}\nn\\
&&-G_5\phi{_{\kappa}}^{\kappa}R_{\mu\nu}
+\square(G_5\phi_{\mu\nu})+(G_5\phi{_{\kappa}}^{\kappa})_{;\nu;\mu}
-2[G_5\phi{_{(\mu}}^{\kappa}]_{;\nu);\kappa}+
\big[(G_5\phi^{\kappa\lambda})_{;\lambda;\kappa}
-\square(G_5\phi{_{\kappa}}^{\kappa})\big]g_{\mu\nu}\Big\}=0
\label{tyg}
\end{eqnarray}
\begin{equation}
{\cal E}_{\phi}=G_{2\phi}+(G_{2X}\phi^{\mu})_{;\mu}-G{_{3\mu}}^{\mu}-G_{3\phi}\Box\phi
+RG_{4\phi}+G_{5\phi}{G}_{\mu\nu}\phi^{\mu\nu}+G_{5}^{\mu\nu}{G}_{\mu\nu}=0\,.
\label{ety}
\end{equation}
Now, again there are terms in (\ref{tyg}) multiplied by $G_{5}$ which contain more than two
derivatives on $\phi$, thus $G_{5}=0$. Equations (\ref{tyg}), (\ref{ety}) become
\begin{eqnarray}
{\cal E}_{\mu\nu}&\!\!=\!\!&
-\frac{1}{2}G_2 g_{\mu\nu}+G_{2X}X_{\mu\nu}
+G_{3(\mu}\phi_{\nu)}-\frac{1}{2}g_{\mu\nu}G_{3\kappa}\phi^{\kappa}
+G_4 G_{\mu\nu}+G_{4\kappa}{^{\kappa}}g_{\mu\nu}-G_{4\mu\nu}=0
\label{vyw}
\end{eqnarray}
\begin{equation}
{\cal E}_{\phi}=G_{2\phi}+(G_{2X}\phi^{\mu})_{;\mu}-G{_{3\mu}}^{\mu}-G_{3\phi}\Box\phi
+RG_{4\phi}=0\,.
\label{myv}
\end{equation}
Equations (\ref{vyw}), (\ref{myv}) are reexpressed as
\begin{equation}
\mathcal{E}_{\mu\nu}=G_{4}G_{\mu\nu}+(G_{2X}-2G_{3\phi}+2G_{4\phi\phi})X_{\mu\nu}-\Big(\frac{1}{2}G_{2}
-G_{3\phi}X+2G_{4\phi\phi}X-G_{4\phi}\Box\phi\Big)g_{\mu\nu}-G_{4\phi}\phi_{\mu\nu}=0
\label{nvr}
\end{equation}
\begin{equation}
\mathcal{E}_{\phi}=G_{2\phi}+RG_{4\phi}+2(G_{3\phi\phi}-G_{2X\phi})X+(G_{2X}-2G_{3\phi})\Box\phi=0\,.
\label{bxs}
\end{equation}
Taking the trace of equation (\ref{nvr}) we obtain the Ricci scalar as
\begin{equation}
G_{4}R=(2G_{3\phi}-6G_{4\phi\phi}-g_{2})X+3G_{4\phi}\Box{\phi}-2\gamma_{2}\,.
\label{okn}
\end{equation}
Substituting this $R$ into (\ref{bxs}) we get
\begin{equation}
G_{4}\gamma_{2\phi}-2\gamma_{2}G_{4\phi}+\big[2G_{4\phi}G_{3\phi}-6G_{4\phi}G_{4\phi\phi}-(g_{2}
G_{4})_{\phi}+2G_{4}G_{3\phi\phi}\big]X+(g_{2}G_{4}-2G_{4}G_{3\phi}+3G_{4\phi}^{2})\Box\phi=0\,.
\label{ugn}
\end{equation}
In order for (\ref{ugn}) to coincide with the wave equation $\Box\phi=0$ of (\ref{lts}) it should
be equivalently
\begin{eqnarray}
&&G_{4}\gamma_{2\phi}-2\gamma_{2}G_{4\phi}=0\label{wnr}\\
&&2G_{4\phi}G_{3\phi}-6G_{4\phi}G_{4\phi\phi}-(g_{2}G_{4})_{\phi}+2G_{4}G_{3\phi\phi}=0\label{juy}\\
&&g_{2}G_{4}-2G_{4}G_{3\phi}+3G_{4\phi}^{2}\neq 0\label{zxd}\,.
\end{eqnarray}
Equation (\ref{wnr}) is immediately integrated to
\begin{equation}
\gamma_{2}=\gamma_{2o}G_{4}^{2}\,,
\label{kiw}
\end{equation}
with $\gamma_{2o}$ integration constant. Then, equation (\ref{nvr}) becomes
\begin{equation}
G_{4}G_{\mu\nu}+\big(g_{2}-2G_{3\phi}+2G_{4\phi\phi}\big)X_{\mu\nu}-\Big(\frac{1}{2}g_{2}
-G_{3\phi}+2G_{4\phi\phi}\Big)Xg_{\mu\nu}-G_{4\phi}\phi_{\mu\nu}-\frac{1}{2}\gamma_{2}g_{\mu\nu}=0\,.
\label{nvk}
\end{equation}
The gravitational equation (\ref{elf}) is written as
\begin{equation}
G_{\mu\nu}+\frac{8\pi}{\phi}\big(2AX_{\mu\nu}+2BXg_{\mu\nu}-C\phi_{\mu\nu}\big)=0\,,
\label{ned}
\end{equation}
where $A,B,C$ satisfy (\ref{jdt}), (\ref{kdw}).
In order for (\ref{nvk}) to coincide with (\ref{ned}) it should be
\begin{eqnarray}
&&g_{2}-2G_{3\phi}+2G_{4\phi\phi}=\frac{16\pi}{\phi}AG_{4}\label{hyb}\\
&&\frac{1}{2}g_{2}-G_{3\phi}+2G_{4\phi\phi}=-\frac{16\pi}{\phi}BG_{4}\label{knm}\\
&&G_{4\phi}=\frac{8\pi}{\phi}CG_{4}\label{lop}\\
&&\gamma_{2}=0\label{hue}\,.
\end{eqnarray}

The action (\ref{horn}) we have resulted up to now is
\begin{equation}
S=\int \!d^{4}x\,\sqrt{-g}\,\big[g_{2}(\phi)X-G_{3}(\phi)\Box\phi+G_{4}(\phi)R\big]\,,
\label{rby}
\end{equation}
where the coefficients $g_{2},G_{3},G_{4}$ obey the system (\ref{juy}), (\ref{zxd}),
(\ref{hyb})-(\ref{lop}). Since the term $G_{3}\Box\phi$ can be converted through an integration
by parts to the term $2G_{3\phi}X$, the action (\ref{rby}) reduces to the simpler form
\begin{equation}
S=\frac{1}{16\pi}\int\!d^{4}x\,\sqrt{-g}\,\big[\textsf{f}(\phi)R-\textsf{h}
(\phi)\phi^{;\mu}\phi_{;\mu}\big]\,,
\label{vtn}
\end{equation}
where $\textsf{f}=16\pi G_{4}$, $\textsf{h}=8\pi(g_{2}-2G_{3\phi})$. Since the system of equations
for $g_{2},G_{3},G_{4}$ is pretty complicated, we will find equivalently from the action (\ref{vtn})
the system of equations that the coefficients $\textsf{f},\textsf{h}$ should satisfy.
From equations (\ref{fhw}), (\ref{lwf}) we obtain
\begin{equation}
G^{\mu}_{\,\,\,\nu}-\frac{1}{\textsf{f}}(\textsf{f}\,''\!+\!\textsf{h})\phi^{;\mu}\phi_{;\nu}
+\frac{1}{\textsf{f}}\Big(\textsf{f}\,''\!+\!\frac{1}{2}\textsf{h}\Big)
\delta^{\mu}_{\,\,\,\nu}\phi^{;\rho}\phi_{;\rho}-
\frac{\textsf{f}\,'}{\textsf{f}}\big(\phi^{;\mu}_{\,\,\,\,;\nu}
\!-\!\delta^{\mu}_{\,\,\,\nu}\Box\phi\big)=0
\label{fhg}
\end{equation}
\begin{equation}
\Big(3\frac{\textsf{f}\,'^{2}}{\textsf{f}}\!+\!2\textsf{h}\Big)\Box\phi
+\Big[\textsf{f}\,'\Big(3\frac{\textsf{f}\,''}{\textsf{f}}\!+\!\frac{\textsf{h}}{\textsf{f}}\Big)
\!+\!\textsf{h}\,'\Big]\phi^{;\mu}\phi_{;\mu}=0\,.
\label{lpf}
\end{equation}
For equation (\ref{lpf}) to coincide with (\ref{lts}) it has to be
\begin{eqnarray}
&&\textsf{f}\,'\Big(3\frac{\textsf{f}\,''}{\textsf{f}}\!+\!\frac{\textsf{h}}{\textsf{f}}\Big)
\!+\!\textsf{h}\,'=0\label{ako}\\
&&3\frac{\textsf{f}\,'^{2}}{\textsf{f}}\!+\!2\textsf{h}\neq 0\label{plu}\,.
\end{eqnarray}
Then, for equation (\ref{fhg}) to coincide with (\ref{elf}) it has to be
\begin{eqnarray}
&&\frac{1}{\textsf{f}}(\textsf{f}\,''\!+\!\textsf{h})=\frac{8\pi}{\phi}A\label{pod}\\
&&\frac{1}{\textsf{f}}\Big(\textsf{f}\,''\!+\!\frac{1}{2}\textsf{h}\Big)=-\frac{8\pi}{\phi}B
\label{vne}\\
&&\frac{\textsf{f}\,'}{\textsf{f}}=\frac{8\pi}{\phi}C\label{lme}\,.
\end{eqnarray}
Substituting $A,B,C$ from (\ref{pod})-(\ref{lme}) into (\ref{jdt}), (\ref{kdw}), and using (\ref{ako}),
we get that (\ref{jdt}), (\ref{kdw}) are identically satisfied. So, the action (\ref{vtn}) is valid
for the system (\ref{elf})-(\ref{lts}) and we remain with equations (\ref{ako}), (\ref{plu}).
The solution of (\ref{ako}) is
\begin{equation}
\textsf{h}(\phi)=\frac{\textsf{s}-3\textsf{f}\,'^{2}(\phi)}{2\textsf{f}(\phi)}\,,
\label{mud}
\end{equation}
where $\textsf{s}$ is integration constant. Equation (\ref{plu}) is satisfied for $s\neq 0$.
Finally, the action of the system (\ref{elf})-(\ref{lts}) takes the form
\begin{equation}
S=\frac{1}{16\pi}\int\!d^{4}x\,\sqrt{-g}\,\Big[\textsf{f}(\phi)R-
\frac{\textsf{s}-3\textsf{f}\,'^{2}(\phi)}{2\textsf{f}(\phi)}
g^{\mu\nu}\phi_{,\mu}\phi_{,\nu}\Big]\,.
\label{vtj}
\end{equation}
The action (\ref{vtj}), as well as the coefficients $A,B,C$ of (\ref{pod})-(\ref{lme}), have been
expressed in terms of one arbitrary function $\textsf{f}(\phi)$. The action (\ref{jgk}) is included in
(\ref{vtj}) if we choose $\textsf{s}=\frac{2\epsilon}{\lambda}$.

\section{A Total Lagrangian}
\label{ejrr}

Now we extend the gravitational action (\ref{jgk}) and consider the total action $S=S_{g}+S_{m}$
including the matter part
\begin{equation}
S_{m}=\int\!d^{4}x\,\sqrt{-g}\,J(\phi)L_{m}\,,
\label{srt}
\end{equation}
where $L_{m}(g_{\kappa\lambda},\Psi)$ is the matter Lagrangian. The symbol $\Psi$ denotes a collection
of extra matter fields and $J(\phi)$ is a function to be
determined. The variation of $S_{m}$ with respect to the metric gives
\begin{equation}
\delta_{g}S_{m}=\frac{1}{2}\int\!d^{4}x\,\sqrt{-g}\,J(\phi)\mathcal{T}^{\mu\nu}
\delta g_{\mu\nu}\,,
\label{skb}
\end{equation}
where the matter energy-momentum tensor is defined as
\begin{equation}
\mathcal{T}^{\mu\nu}=\frac{2}{\sqrt{-g}}\frac{\delta(\sqrt{-g}\,L_{m})}{\delta g_{\mu\nu}}
\label{bne}
\end{equation}
in the Jordan frame we are working with, where $\phi$ has its distinctive role in (\ref{elp}) and
$L_{m}$ is multiplied by the non-trivial factor $J(\phi)$ in (\ref{srt}).
The way $\mathcal{T}^{\mu\nu}$ arises from $S_{m}$ is such that the gravitational equation
(\ref{elp}) can be obtained. The variation of the total action is
\begin{equation}
\delta_{g}S=-\frac{1}{16\pi}\int\!d^{4}x\,\sqrt{-g}\,f\Big(\mathcal{E}^{\mu\nu}\!-\!8\pi
\frac{J}{f}\mathcal{T}^{\mu\nu}\Big)\delta g_{\mu\nu}\,,
\label{ccd}
\end{equation}
and therefore the total gravitational equation of motion is
\begin{equation}
\mathcal{E}^{\mu}_{\,\,\,\nu}\!-\!8\pi\frac{J}{f}\mathcal{T}^{\mu}_{\,\,\,\,\nu}=0\,.
\label{kcb}
\end{equation}
In order for (\ref{kcb}) to coincide with equation (\ref{elp}), it should be $J=\frac{f}{\phi}$, thus
\begin{equation}
J(\phi)=\frac{1}{\sqrt{8\pi}}\,\frac{\sqrt{|\nu\!+\!8\pi\phi^{2}|}}{|\phi|}\,.
\label{xcb}
\end{equation}
Note that for $\nu=0$, it becomes $J=1$. Therefore, a candidate
total action for the complete Brans-Dicke theory (\ref{elp})-(\ref{idj}) is
\begin{equation}
S=\frac{\eta}{2(8\pi)^{3/2}}\int\!d^{4}x\,\sqrt{-g}\,
\Big[\sqrt{|\nu\!+\!8\pi\phi^{2}|}\,R-\frac{8\pi}{\lambda}\,
\frac{\nu\!+\!4\pi(2\!-\!3\lambda)\phi^{2}}{|\nu\!+\!8\pi\phi^{2}|^{3/2}}g^{\mu\nu}\phi_{,\mu}\phi_{,\nu}
+16\pi\frac{\sqrt{|\nu\!+\!8\pi\phi^{2}|}}{\phi}L_{m}(g_{\kappa\lambda},\Psi)\Big]\,.
\label{msi}
\end{equation}
Notice that due to the interaction term in the conservation equation (\ref{idj}) with $\nu\neq 0$,
the matter Lagrangian is non-minimally coupled even in the Jordan frame.

We have not yet finished having derived
the equation (\ref{elp}) since we have not discussed the derivation of equations (\ref{lrs}) and
(\ref{idj}). As for (\ref{idj}), if (\ref{lrs}) has been derived, then (\ref{idj})
is the consistency condition in order for the Bianchi identities to be satisfied which have
indeed been verified in \cite{Kofinas:2015nwa}. Thus, it remains to see if equation (\ref{lrs})
is obtained under variation of (\ref{msi}) with respect to $\phi$. It is evident that a
variation of $S_{m}$ with respect to $\phi$ will produce a factor $L_{m}$ which cannot
be canceled by any other equation. More precisely, extending equation (\ref{qwk}), we obtain
\begin{equation}
\delta_{\phi}S=\frac{1}{16\pi}\int\!d^{4}x\,\sqrt{-g}\,\big(f'R+h'\phi^{;\mu}\phi_{;\mu}
+2h\Box\phi+16\pi J' L_{m}\big)\delta\phi\,.
\label{jrg}
\end{equation}
The scalar field equation is thus
\begin{equation}
f'R+h'\phi^{;\mu}\phi_{;\mu}+2h\Box\phi+16\pi J' L_{m}=0\,.
\label{jwm}
\end{equation}
The trace of equation (\ref{kcb}) gives
\begin{equation}
R=\frac{1}{f}(3f''\!+\!h)\phi^{;\mu}\phi_{;\mu}+3\frac{f'}{f}\Box\phi-8\pi
\frac{J}{f}\mathcal{T}\,.
\label{wjf}
\end{equation}
From (\ref{jwm}), (\ref{wjf}) we obtain
\begin{equation}
\Big(3\frac{f'^{2}}{f}\!+\!2h\Big)\Box\phi+\Big[f'\Big(3\frac{f''}{f}\!+\!\frac{h}{f}\Big)\!+\!h'\Big]
\phi^{;\mu}\phi_{;\mu}
-8\pi\frac{f'}{f}J\mathcal{T}+16\pi J' L_{m}=0\,.
\label{kld}
\end{equation}
Simplifying this equation, we finally get
\begin{equation}
\Box\phi=4\pi\lambda\mathcal{T}+\frac{\lambda\nu}{\phi^{2}}L_{m}\,.
\label{jer}
\end{equation}
Therefore, the field equation (\ref{lrs}) is obtained only if on-shell the numerical value of the
matter Lagrangian $L_{m}$ is zero. For example, for a relativistic perfect fluid,
an action functional has been constructed \cite{Schutz:1970my} where the matter Lagrangian
is proportional to the pressure.
Moreover, in \cite{Brown:1992kc} the on-shell Lagrangian, i.e. the value of the
Lagrangian when the equations of motion hold, is again the pressure, thus for pressureless dust this
on-shell value vanishes. Of course, an energy-momentum tensor is normally defined and enters the
field equations, because this is computed off-shell.

The result of this section is that in the case of matter, we have found an action functional
of the form (\ref{msi}) for the complete Brans-Dicke theory studied,
only for particular matter Lagrangians, those which vanish on-shell. This is still meaningful, although
of restricted applicability. This result, however, does not mean that we have shown that
an action principle does not exist for arbitrary matter Lagrangians.
It is an option that actions of a different, more complicated form than (\ref{msi}),
could in principle exist and provide the full set of field equations with any matter content.

A notice of caution should be added at this point.
It is true that if equation (\ref{idj}) had been derived, then equation (\ref{lrs}) would arise from
the satisfaction of the Bianchi identities. Note first that the conservation equation (\ref{idj})
is also written as
\begin{equation}
\Big(\frac{\sqrt{|\nu\!+\!8\pi\phi^{2}|}}{\phi}\mathcal{T}^{\mu}_{\,\,\,\,\nu}\Big)_{;\mu}=0\,.
\label{xdg}
\end{equation}
Comparing this equation with the last term in the action (\ref{msi}), the one containing $L_{m}$,
one could be tempted to apply the standard diffeomorphism invariance argument for this part
of the action, and obtain immediately equation (\ref{xdg}). However, this is not correct
because of the non-minimal coupling of $L_{m}$ with $\phi$.

The action (\ref{msi}), whenever applied, can be cast into a canonical form where the Einstein-Hilbert
term is only minimally coupled. Of course, the transformations following are also valid in the
vacuum case. We perform a new conformal transformation
\begin{equation}
\tilde{g}_{\mu\nu}=\tilde{\Omega}^{2}(\phi)g_{\mu\nu}\,\,\,\,\,\,,\,\,\,\,\,\,
\tilde{\Omega}=\Big(\frac{|\nu\!+\!8\pi\phi^{2}|}{8\pi}\Big)^{\frac{1}{4}}\,.
\label{mnb}
\end{equation}
If $\tilde{R}$, $\tilde{\Box}$ correspond to $\tilde{g}_{\mu\nu}$ and
$\tilde{\omega}=\tilde{\Omega}^{-1}$, the total action $S$ defined from (\ref{ker}),
(\ref{srt}), using (\ref{mje}), takes the form
\begin{equation}
S=\frac{1}{16\pi}\int\!d^{4}x\,\sqrt{-\tilde{g}}\,f\tilde{\omega}^{2}\Big{\{}\tilde{R}+\Big[6
\frac{\tilde{\omega}'}{\tilde{\omega}}\,\frac{(f\tilde{\omega}^{2})'}{f\tilde{\omega}^{2}}
\!-\!6\frac{\tilde{\omega}'^{2}}{\tilde{\omega}^{2}}\!-\!\frac{h}{f}\Big]
\tilde{g}^{\mu\nu}\phi_{,\mu}\phi_{,\nu}\Big{\}}
+\int\!d^{4}x\,\sqrt{-\tilde{g}}\,\frac{f\tilde{\omega}^{4}}{\phi}\,L_{m}(\tilde{\omega}^{2}
\tilde{g}_{\kappa\lambda},\Psi)\,.\label{fno}
\end{equation}
Using that $f\tilde{\omega}^{2}=\eta$, $\frac{\tilde{\omega}'}{\tilde{\omega}}=\frac{1}{4}
\frac{(\tilde{\omega}^{4})'}{\tilde{\omega}^{4}}=-\frac{4\pi\phi}{\nu+8\pi\phi^{2}}$, we find
an action with the Einstein-Hilbert term without non-minimal coupling, but with a non-standard
kinetic term
\begin{equation}
S=\frac{\eta}{16\pi}\int\!d^{4}x\,\sqrt{-\tilde{g}}\,\Big[\tilde{R}-\frac{8\pi}{\lambda
(\nu\!+\!8\pi\phi^{2})}\tilde{g}^{\mu\nu}\phi_{,\mu}\phi_{,\nu}+\frac{2(8\pi)^{\frac{3}{2}}}
{\phi\sqrt{|\nu\!+\!8\pi\phi^{2}|}}L_{m}(\tilde{\omega}^{2}\tilde{g}_{\kappa\lambda},\Psi)\Big]\,.
\label{nty}
\end{equation}
In order to make this kinetic term canonical, we introduce a new scalar field $\sigma(x)$, instead of
$\phi(x)$, by
\begin{equation}
\frac{d\phi}{d\sigma}=\sqrt{\frac{|\lambda|}{16\pi}}\,\sqrt{|\nu\!+\!8\pi\phi^{2}|}\,.
\label{jcy}
\end{equation}
The kinetic term in (\ref{nty}) inside the bracket is re-expressed as
$-\frac{1}{2}\epsilon\epsilon_{\lambda}\tilde{g}^{\mu\nu}\sigma_{,\mu}\sigma_{,\nu}$, where
$\epsilon_{\lambda}$ is the sign of $\lambda$.
This is a canonical kinetic term, so the field $\sigma$ behaves as a usual scalar field in the new
conformal frame. If
$\epsilon\epsilon_{\lambda}>0$, then $\sigma$ is a normal field with positive energy.
This is achieved even if the kinetic term in (\ref{msi}) is positive. Although $\phi$ is
``apparently'' a ghost in this case since it has the wrong sign, however, $\sigma$ is not a ghost.
On the opposite, if $\epsilon\epsilon_{\lambda}<0$, then $\sigma$ is a ghost.

For $\epsilon>0$, equation (\ref{jcy}) is integrated to
\begin{equation}
\sigma-\sigma_{0}=
\sqrt{\frac{2}{|\lambda|}}\,\ln{\Big|4\pi\phi\!+\!\sqrt{2\pi}\sqrt{\nu\!+\!8\pi\phi^{2}}\Big|}\,,
\label{kil}
\end{equation}
where $\sigma_{0}$ is integration constant. Redefining $\sigma_{0}$, we can rewrite (\ref{kil}) in the
form
\begin{equation}
\sigma=\sqrt{\frac{2}{|\lambda|}}\,\ln{\Bigg|\frac{4\pi\phi\!+\!\sqrt{2\pi}\sqrt{\nu\!+\!8\pi\phi^{2}}}
{4\pi\phi_{0}\!+\!\sqrt{2\pi}\sqrt{\nu\!+\!8\pi\phi_{0}^{2}}}\Bigg|}\,,
\label{dio}
\end{equation}
where $\phi_{0}$ is integration constant. In the case that the scalar field $\phi$ is constant with value
$\phi=\phi_{0}$, it will be $\sigma=0$. In this case there is no need for a conformal transformation
and we may then set $\tilde{\Omega}=1$. Thus, $\phi_{0}^{2}=1-\frac{\nu}{8\pi}$ and the integration
constant $\phi_{0}$ has been determined (for $\nu>0$ it should be $\nu<8\pi$). We redefine $\sigma$
by absorbing the translational integration constant $\sigma_{0}$ of (\ref{kil}) into $\sigma$.
Equation (\ref{kil}) can be inverted giving
\begin{equation}
\phi=\frac{s}{8\pi}\Big(e^{\sqrt{\frac{|\lambda|}{2}}\,\sigma}\!-\!2\pi\nu
e^{-\sqrt{\frac{|\lambda|}{2}}\,\sigma}\Big)\,,
\label{nui}
\end{equation}
where $s=\text{sgn}(4\pi\phi\!+\!\sqrt{2\pi}\sqrt{\nu+8\pi\phi^{2}})=
\text{sgn}\Big(e^{\sqrt{\frac{|\lambda|}{2}}\,\sigma}\!+\!2\pi\nu
e^{-\sqrt{\frac{|\lambda|}{2}}\,\sigma}\Big)$. The action (\ref{nty}) becomes
\begin{equation}
S=\frac{\eta}{16\pi}\int\!d^{4}x\,\sqrt{-\tilde{g}}\,\Big[\tilde{R}-
\frac{1}{2}\epsilon_{\lambda}\tilde{g}^{\mu\nu}\sigma_{,\mu}
\sigma_{,\nu}+\frac{2\eta(8\pi)^{3}}{\big|e^{\sqrt{2|\lambda|}\,\sigma}\!-\!4\pi^{2}\nu^{2}
e^{-\sqrt{2|\lambda|}\,\sigma}\big|}L_{m}(\tilde{\omega}^{2}\tilde{g}_{\kappa\lambda},\Psi)\Big]\,,
\label{vtl}
\end{equation}
where
\begin{equation}
\tilde{\omega}^{2}=\frac{8\pi}{\Big|e^{\sqrt{\frac{|\lambda|}{2}}\,\sigma}\!+\!2\pi\nu
e^{-\sqrt{\frac{|\lambda|}{2}}\,\sigma}\Big|}\,.
\label{guh}
\end{equation}
For the physically more interesting case with $\phi>0$, both absolute values in (\ref{vtl}),
(\ref{guh}) disappear.
The Lagrangian (\ref{vtl}) refers to the Einstein frame where the gravitational coupling is a true
constant. In order for $\sigma$ not to be a ghost, it should be $\lambda>0$.
In the Brans-Dicke limit $\nu=0$ the coupling to $L_{m}$ is a simple exponential function
of $\sigma$.

For $\epsilon<0$, equation (\ref{jcy}) is integrated to
\begin{equation}
\sigma-\sigma_{0}=\sqrt{\frac{2}{|\lambda|}}\,\arcsin{\Big(\sqrt{\frac{8\pi}{|\nu|}}\,\phi\Big)}\,,
\label{pou}
\end{equation}
where $\sigma_{0}$ is integration constant. Redefining $\sigma_{0}$, we can write (\ref{pou})
in the form
\begin{equation}
\sigma=\sqrt{\frac{2}{|\lambda|}}\,
\Big[\arcsin{\Big(\sqrt{\frac{8\pi}{|\nu|}}\,\phi\Big)}
\!-\!\arcsin{\Big(\sqrt{\frac{8\pi}{|\nu|}}\,\phi_{0}\Big)}\Big]\,,
\label{poi}
\end{equation}
where $\phi_{0}$ is integration constant. Again for $\phi=\phi_{0}$ it is $\sigma=0$, and setting
$\tilde{\Omega}=1$ we get the condition $\phi_{0}^{2}=\frac{|\nu|}{8\pi}-1$ (it should
be $|\nu|>8\pi$). We redefine $\sigma$ by absorbing the translational integration constant
$\sigma_{0}$ of (\ref{pou}) into $\sigma$, and then
$-\frac{\pi}{2}<\sqrt{\frac{|\lambda|}{2}}\,\sigma<\frac{\pi}{2}$.
Equation (\ref{pou}) can be inverted giving
\begin{equation}
\phi=\sqrt{\frac{|\nu|}{8\pi}}\,\sin\Big(\sqrt{\frac{|\lambda|}{2}}\,\sigma\Big)\,.
\label{yun}
\end{equation}
The action (\ref{nty}) becomes
\begin{equation}
S=\frac{\eta}{16\pi}\int\!d^{4}x\,\sqrt{-\tilde{g}}\,\Big[\tilde{R}+
\frac{1}{2}\epsilon_{\lambda}\tilde{g}^{\mu\nu}\sigma_{,\mu}
\sigma_{,\nu}+\frac{4(8\pi)^{2}}{|\nu|\,\sin\big(\sqrt{2|\lambda|}\,\sigma\big)}
L_{m}(\tilde{\omega}^{2}\tilde{g}_{\kappa\lambda},\Psi)\Big]\,,
\label{vtx}
\end{equation}
where
\begin{equation}
\tilde{\omega}^{2}=\frac{\sqrt{8\pi}}{\sqrt{|\nu|}\,\cos\Big(\sqrt{\frac{|\lambda|}{2}}\,\sigma\Big)}\,.
\label{gug}
\end{equation}
In order for $\sigma$ not to be a ghost, it should be $\lambda<0$.

\section{Conclusions}
\label{Conclusions}

Relieving the standard exact conservation of matter, but still preserving the simple wave equation
of motion for the scalar field sourced by the trace of the matter energy-momentum tensor,
it was recently found the most general completion of Brans-Dicke
theory. This class of theories contains three interaction terms in the non-conservation
equation of matter and is parametrized by arbitrary functions of the scalar field. Keeping a single
interaction term each time to express the energy exchange between the scalar field and ordinary matter,
three uniquely defined theories arise from consistency, which form the prominent and natural complete
Brans-Dicke theories.

Here, for the first such theory, its vacuum part, which arises as the
zero-matter limit, is studied. The Lagrangian of this vacuum theory is found in the so called
Jordan frame, where the scalar field plays the role of the inverse gravitational parameter in the
field equations. In this frame, a symmetry transformation of the vacuum action is also found,
which consists of a conformal transformation of the metric together with a redefinition of the
scalar field. Since the general family of vacuum theories is not exhausted by the above vacuum
theory but contains a free function of the scalar field, the action of this family is found which
is also parametrized by an arbitrary function and forms a subclass of the Horndeski theories.

As for the corresponding total theory with matter, we have not been able to answer the question if the
complete Brans-Dicke theory studied here, with a general matter energy-momentum tensor, arises or
not from a Lagrangian. We have only answered this question partially in the case that the matter
Lagrangian vanishes on-shell, as for example happens in the case of pressureless dust.
Due to the interaction term in the conservation equation, the matter Lagrangian is non-minimally
coupled even in the Jordan frame. In the Einstein frame where the Einstein-Hilbert term is minimally
coupled, two forms of this total action have been found, one with a non-canonical kinetic term and
one with a canonical kinetic term, with the matter Lagrangian still being non-minimally coupled,
while these forms of the actions still make sense in the vacuum limit.

\[ \]
{{\bf Acknowledgements}} We wish to thank K. Ntrekis and E.
Papantonopoulos for useful discussions. M.T. is supported by TUBITAK
2216 fellowship under the application number 1059B161500790.

%\appendix

%\section{Geometric Components} \label{geometric components}

%%%%%%%%%%%%%%%%%%%%%%%%%%%% BIBLIOGRAPHY %%%%%%%%%%%%%%%%%%%%%%%%%%%%%%%%%%%%%%

\end{document}